\documentclass[aps,prd,twocolumn,groupedaddress,amssymb,eqsecnum,showpacs,
nofootinbib]{revtex4}
\usepackage{graphicx}
\usepackage{bm}

\begin{document}

\preprint{CMU-HEP-03-05}

\title{The enhancement of inflaton loops in an $\alpha$-vacuum}

\author{Hael Collins}
\email{hcollins@andrew.cmu.edu}
\author{Matthew R.~Martin}
\email{mmartin@cmu.edu}
\affiliation{Department of Physics, Carnegie Mellon University, 
Pittsburgh PA\ \ 15213}

\date{\today}

\begin{abstract}
While inflaton loops in the Euclidean vacuum generally have a negligible contribution to the power spectrum, loop effects can be substantially larger when the inflaton is in a non-thermal vacuum state.  As an example, we show that in a truncated $\alpha$-vacuum these loop effects are enhanced by the ratio of the Planck scale to the Hubble scale during inflation.  The details of the inflationary models determine whether the coupling constants suppress the loop corrections relative to the tree level result.
\end{abstract}


\maketitle

\section{Introduction}\label{intro}

The extremely rapid expansion which occurs during inflation naturally connects quantum fluctuations on the smallest scales to large astrophysical distances today.  If inflation lasted for a sufficiently long time, the scales associated with cosmological structure could have been generated when they were nearly the Planck scale or smaller.  Observations of the cosmic microwave background (CMB) radiation, such as the recent results of the Wilkinson Microwave Anisotropy Probe \cite{wmap}, are reaching a level of precision which are able not only to confirm the generic predictions of inflation, but also should be capable of distinguishing various specific models for generating inflation.  Physics at energies above the Hubble scale during inflation will also be generically imprinted on the CMB and could, if the Hubble parameter is sufficiently large, be observable to future experiments such as PLANCK \cite{planck}.

Within a general inflationary model, the spectrum of density fluctuations of a scalar field about its vacuum state become the temperature variations observed in the CMB radiation.  In order to understand what the detailed form of the power spectrum reveals about short distance physics, we must estimate the size of the deviations from the standard result, which are produced by any new physics.  The standard calculation of the power spectrum of the density fluctuations assumes that the universe is in a particular vacuum state known as the Euclidean, Bunch-Davies or thermal vacuum \cite{bunch}.  Even in the Euclidean vacuum, any new physics with a mass scale $\Lambda_{\rm phys}$ above the Hubble constant during inflation, $H$, will leave an imprint on the CMB although the deviation will be suppressed by $H^2/\Lambda_{\rm phys}^2$ \cite{kaloper}.  However, when the universe is not in the Euclidean vacuum during inflation, the effects of the short distance physics can be substantially larger, scaling as $H/\Lambda_{\rm phys}$ \cite{rich,ulf,kempf,gary,garygeneric,lowe,brandenberger,ss}, and could be observable if the scale of inflation is sufficiently high.  These deviations occur in tree-level calculations, but loop effects could also be important since they are enhanced compared with their contribution in the Euclidean vacuum.  

The motivation for choosing a vacuum other than the Euclidean vacuum has many origins.  If we look at a length scale of cosmological relevance today, during inflation this scale was exponentially smaller being further blue-shifted the further we look back.  If inflation lasted sufficiently long, more than the necessary 60 $e$-folds needed to solve the flatness and homogeneity problems, scales that we observe today were of Planckian size at some point during inflation.  At such short distances, the scalar field used to generate the fluctuations may be modified by strong gravity effects.  Therefore, without an understanding of field theory at Planckian scales, we must impose some matching condition on the low energy theory at this scale.  Further, without some mechanism for thermalization near the Planck scale, such as that which solves a similar transplanckian problem for black holes \cite{blackhole}, there is no compelling reason that these matching conditions select the thermal, Euclidean vacuum.  In fact, some amount of non-thermality could be natural in inflation \cite{ulftherm, chung}.

In the case of a scalar field in a de Sitter space-time---exponential inflation---a large family of vacua exists which are invariant under the full group of de Sitter symmetries \cite{mottola,allen}.  These vacua can be labeled by a complex parameter $\alpha$ and are called the $\alpha$-vacua and include the Euclidean vacuum as a special case.  These vacua are distinguished by their differing behavior at short distances and can be used as examples of different states which could emerge when matched onto the unknown physics above the Planck scale.  Their relatively simple structure makes them a suitable setting in which to study the perturbative corrections generated when the inflaton field has self-interactions.

An interacting quantum field theory in a pure $\alpha$-vacuum seems to suffer from many pathologies.\footnote{Even without interactions, the pure $\alpha$-vacuum may not be sensible \cite{jackiw,kklss}.}  Perturbative corrections either produce nonrenormalizable ultraviolet divergences \cite{fate}, pinched singularities \cite{einhorn} or require non-local counterterms \cite{banks}.\footnote{These pathologies might also signal that the usual rules for quantizing field theories must be generalized if we wish to study interactions in an $\alpha$-vacuum \cite{goldstein} or that we ought to study generalizations of the $\alpha$-vacuum which can be renormalized \cite{squeezed}.  Any generalization however should reduce to the usual methods in the Euclidean limit.}  However, Danielsson \cite{ulf} has emphasized that to be physically relevant for the transplanckian problem in inflation, the modes in the $\alpha$-vacuum should not be defined to arbitrarily high energies, but only up to a cutoff of the order of the Planck scale, $M_{\rm pl}$.  Imposing a cutoff on the theory removes the divergences since the loop momenta, which led to the nonrenormalizable terms, no longer can be arbitrarily large.  Note that even for a finite cutoff, such terms, although finite, still cannot be absorbed by a local counterterm.

Even in such a `truncated $\alpha$-vacuum', perturbative corrections could be substantially larger for a general $\alpha$ than for the Euclidean limit---the loop corrections may no longer diverge, but they are still enhanced by powers of the cutoff scale.  In calculating the shape of the power spectrum in a truncated $\alpha$-vacuum it is therefore important to assess the size of these perturbative corrections since they could be comparable to other non-thermal tree level effects.  In particular, we shall show that loop corrections for a theory with a cubic interaction with a coupling of $\lambda$ is only suppressed by a factor of $(\lambda/H)^2 (\Lambda_{\rm phys}/H)$ relative to the Euclidean tree-level result.  The factor $\Lambda_{\rm phys}$ encodes the sensitivity to new physics, $\Lambda_{\rm phys}\sim M_{\rm pl}$.  The $\Lambda_{\rm phys}/H$ factor depends only on the structure of the loop corrections in the $\alpha$-vacuum while the initial $(\lambda/H)^2$ factor depends on the details of the inflationary model.  For example, in chaotic inflation, the small size of the coupling is sufficient to overcome the loop enhancement; however, in hybrid models the couplings can be significantly larger so that the total effect of the loop corrections can be comparable to the tree level result.

In the next section we review the invariant vacua of de Sitter space and introduce the truncated $\alpha$-vacuum we shall use.  Our calculation of the one-loop corrections to the power spectrum in the truncated $\alpha$-vacuum appears in Section~\ref{power}.  The perturbation theory is developed using the Schwinger-Keldysh method for studying the finite time evolution of a quantum field theory from an initial state.  Section~\ref{potential} describes constraints on the size of the coupling of the inflaton and we present our conclusions in Section~\ref{conclude}.

\section{A truncated $\alpha$-vacuum}\label{truncated}

The theory of a free scalar field in a de Sitter space-time has a one parameter family of vacua labeled by the complex parameter $\alpha$.  Each of these vacua is invariant under the symmetries of de Sitter space and this property is most easily demonstrated by showing that the Wightman two-point function in an $\alpha$-vacuum depends only on the de Sitter invariant distance between the points \cite{mottola,allen,bousso}.  With respect to the space-time symmetries then, any of these $\alpha$-vacua provides an acceptable choice for the vacuum state.  In this section we shall first review the properties of the true $\alpha$-vacuum before introducing the phenomenologically more realistic truncated $\alpha$-vacuum.  Our notation follows \cite{fate} where perturbation theory in an $\alpha$-vacuum is developed more fully.

At distances shorter than the inherent curvature length associated with de Sitter space, the space-time appears approximately flat.  In this limit, it should be possible to apply the same prescription for defining positive and negative frequency modes as in Minkowski space.  The unique vacuum which matches with the Minkowski vacuum in this limit is the Euclidean, or Bunch-Davies vacuum.  In addition to reducing to the flat space vacuum at high energies, this vacuum also is thermal---an Unruh detector placed in this vacuum satisfies the principle of detailed balance as though the background is at the de Sitter temperature, $T_{\rm dS}=H/2\pi$ \cite{birrelldavies}.  For these reasons, the Euclidean vacuum is frequently assumed to be the correct choice when calculating the power spectrum of the primordial fluctuations which seed the temperature fluctuations that appear in the CMB radiation.

A convenient choice of coordinates for studying de Sitter space is provided by conformally flat coordinates, 
\begin{equation}
ds^2 = {d\eta^2 - d\vec x^2\over H^2\eta^2} , 
\qquad
\eta\in [-\infty,0] . 
\label{metric}
\end{equation}
These coordinates are simply related to the standard coordinates used in inflation,
\begin{equation}
ds^2 = dt^2 - e^{2Ht}\, d\vec x^2 , 
\label{inflatemetric}
\end{equation}
through $\eta = - H^{-1}e^{-Ht}$.  The Hubble constant is related to the cosmological constant which is equal to $6H^2$.  

In a model in which the density perturbations are seeded by the fluctuations of a scalar field, the inflaton is divided into terms describing respectively the spatially homogeneous zero mode, $\phi(\eta)$, and a term for the fluctuations, $\Phi(\eta,\vec x)$, 
\begin{equation}
\hat\Phi(\eta,\vec x) = \phi(\eta) + \Phi(\eta,\vec x) .
\label{inflaton}
\end{equation}
$\phi(\eta)$ is the mode which drives inflation.  The expansion of a free scalar field with respect to the Euclidean vacuum is then given by 
\begin{equation}
\Phi(\eta,\vec x) = \int {d^3\vec k\over (2\pi)^3}\, 
\left[ U_k^E(\eta) e^{i\vec k\cdot \vec x} a_{\vec k}^E 
+ U_k^{E*}(\eta) e^{-i\vec k\cdot \vec x} a_{\vec k}^{\dagger E} \right]
\label{opexpand}
\end{equation}
where the operator $a_{\vec k}^E$ annihilates the Euclidean vacuum, $|E\rangle$.  The Euclidean mode functions $U_k^E(\eta)$ are solutions to the Klein-Gordon equation which in conformally flat coordinates is 
\begin{equation}
\left[ \eta^2 {\partial^2\over\partial\eta^2} - 2\eta {\partial\over\partial\eta} + \eta^2k^2 + {m^2\over H^2} \right] 
U_k^E(\eta) = 0 . 
\label{KleinGordon}
\end{equation}
Since the Euclidean modes should become those of the Minkowski vacuum at short distances, or as $H\to 0$, they are fixed to be  
\begin{equation}
U_k^E(\eta) = {\sqrt{\pi}\over 2} H\eta^{3/2} H_\nu^{(2)}(k\eta) 
\label{Emodes}
\end{equation}
where $H_\nu^{(2)}(k\eta)$ is the Hankel function and where
\begin{equation}
\nu \equiv \sqrt{\textstyle{ {9\over 4} - {m^2\over H^2} }} .
\label{nudef}
\end{equation}
To obtain the correct tree level prediction for the CMB power spectrum, we shall need the effective mass of the inflaton, $m$, to be small compared with the Hubble scale so we shall frequently consider the limit of a massless, minimally coupled field although it is important to remember that a small but finite mass is present to avoid infrared divergences in the theory.  In the limit $m\to 0$, the Euclidean mode function simplifies to 
\begin{equation}
U_k^E(\eta) = {iH\over k\sqrt{2k}} (1+ik\eta) e^{-ik\eta} . 
\label{Emodemmc}
\end{equation}

The mode functions for the $\alpha$-vacua can be obtained from those for the Euclidean vacuum through
\begin{equation}
U_k^\alpha(\eta) = N_\alpha \left[ U_k^E(\eta) + e^\alpha U_k^{E*}(\eta) \right] 
\label{Amodes}
\end{equation}
where
\begin{equation}
N_\alpha = \left( 1 - e^{\alpha+\alpha^*} \right)^{-1/2}
\label{Nalphadef}
\end{equation}
since the operator that annihilates the $\alpha$-vacuum, $|\alpha\rangle$, is a Bogolubov transformation of the Euclidean vacuum creation and annihilation operators.  Here ${\rm Re}\, \alpha < 0$ and note that we recover the Euclidean vacuum in the limit, $\alpha\to -\infty$.

The mode functions that appear in Eq.~(\ref{Emodes}) and Eq.~(\ref{Amodes}) are valid for a free scalar field in a classical de Sitter background.  At sufficiently high energies $\Lambda_{\rm phys}\sim M_{\rm pl}$, we expect that gravity becomes strongly interacting and there is no reason to assume that the mode functions for the full theory will continue to satisfy the Klein-Gordon equation (\ref{KleinGordon}).  In order to study the size of loop effects at energies where gravity is weakly interacting, we shall consider a simple scenario in which the strongly interacting theory rapidly damps the mode functions at high energies while at lower scales the mode functions are simply those of the free theory,
\begin{equation}
U_k(\eta) = \cases{U_k^\alpha(\eta) &for $k\le \Lambda$ \cr
                   0                &for $k >  \Lambda$ \cr} . 
\label{fullmodes}
\end{equation}
Note that $k$ is the comoving and not the physical momentum, so the cutoff $\Lambda$ depends on the conformal time, $\Lambda = \Lambda_{\rm phys}/(-H\eta)$.  Eq.~(\ref{fullmodes}) defines a `truncated $\alpha$-vacuum.'

Although we shall use this truncated $\alpha$-vacuum to estimate the size of loop corrections to the power spectrum, the most general case could include some dependence on the momentum as well when we match onto the short distance physics,
\begin{equation}
U_k^{\rm gen}(\eta) = {A_k\over k\sqrt{2k}}\, (1+ik\eta) e^{-ik\eta} 
                    + {B_k\over k\sqrt{2k}}\, (1-ik\eta) e^{ik\eta} 
\label{genmodes}
\end{equation}
for $k<\Lambda$.  $A_k$ and $B_k$ are not independent since they are related by the normalization of the state.  The final degree of freedom for the mode is then fixed by some assumption about the matching of the mode to the high energy theory, at $k=\Lambda$ \cite{ulf,kempf,gary,garygeneric,lowe,brandenberger,ss}.  Depending upon the matching condition, the coefficients $A_k$ and $B_k$ may introduce some additional $k$-dependence which will appear in the power spectrum.  Here we shall study the case when the high energy theory matches onto an $\alpha$-vacuum at $k=\Lambda$.  Although this case will not introduce any new $k$-dependence into the power spectrum, it will allow us to estimate the size of perturbative corrections when the universe is not in the Euclidean vacuum during inflation.  Our results then will indicate how large similar loop effects from other $k$-dependent vacua could be.

\section{The power spectrum}\label{power}

In the standard inflationary picture, the seeds of the large scale structure are provided by the quantum fluctuations of a scalar field during inflation.  This scalar field is given by a linear combination of the fluctuations of the inflaton, $\Phi(\eta,\vec x)$, and the scalar component of the metric fluctuations \cite{mukhanov,recentpert}.  In the limit approaching purely exponential inflation, the inflaton fluctuations become the dominant component of this linear combination so in this section we shall neglect the contribution of the metric fluctuations.  The power spectrum of density fluctuations, which produce the CMB temperature fluctuations and which eventually become the large scale structure, in this limit is proportional to the power spectrum of the scalar field, ${\cal P}^\alpha(\eta,k)$, defined by 
\begin{equation}
\langle\alpha |\, \Phi(\eta,\vec x) \Phi(\eta,\vec y)\, | \alpha\rangle
\equiv \int {d^3\vec k\over (2\pi)^3}\, e^{i\vec k\cdot (\vec x-\vec y)} \left[ {2\pi^2\over k^3} {\cal P}^\alpha(\eta,k) \right]
\label{powerdef}
\end{equation}
up to corrections suppressed by the slow roll parameters which are small as we approach a purely de Sitter limit.

At tree level, substituting in the mode expansion of Eq.~(\ref{opexpand}) and using massless Euclidean mode functions in Eq.~(\ref{Emodemmc}), the power spectrum for the Euclidean vacuum is 
\begin{equation}
{\cal P}^E(\eta,k) = {H^2\over 4\pi^2} (1+k^2\eta^2) . 
\label{Epower}
\end{equation}
When a mode has been redshifted well outside the horizon during inflation, $k\eta\ll 1$, the power spectrum becomes flat.  In a general $\alpha$-vacuum, the power spectrum to leading order in $k\eta$ is also flat,
\begin{equation}
{\cal P}^\alpha(\eta,k) = {H^2\over 4\pi^2} 
N_\alpha^2 |1 - e^\alpha|^2 \left[ 1 + {\cal O}(k^2\eta^2) \right] . 
\label{Apower}
\end{equation}
As the $\alpha$-dependent prefactor is not large, it is observationally difficult to distinguish from other cosmological parameters unless $\alpha$ were to have some $k$-dependence \cite{ulf,kempf,gary,garygeneric,lowe,brandenberger,ss}.  

To study the perturbative corrections to the power spectrum in a truncated $\alpha$-vacuum, we shall consider a theory with a cubic interaction.  This example provides a simple setting since the first non-trivial corrections to the two-point function (\ref{powerdef}) appear already at one loop order.

One of the difficulties in formulating perturbation theory in a de Sitter background is the lack of a well-defined $S$-matrix.  Therefore, we should apply a quantization procedure that evolves a matrix element over a finite conformal time interval rather than one that evaluates the matrix element between asymptotic `in' and `out ' states.  This approach also allows us to avoid the transplanckian problem since we can choose the initial state as that given at the matching scale $\Lambda$.  If we had attempted to follow an `in' state back earlier, then we would eventually need to evaluate the state when the physical scales relevant today would have been blue-shifted above the Planck scale.  

The closed time contour formalism developed by Schwinger, Keldysh and others \cite{schwinger,keldysh,kt} describes a perturbative approach for solving the evolution of a matrix element over a finite time interval.  Unlike the usual $S$-matrix calculation which essentially requires only a single insertion of the time evolution operator, in the Schwinger-Keldysh approach both the $\langle \alpha |$ and the $|\alpha\rangle$ states are evolved from a known initial state at $\eta_0$ to a finite time later, $\eta$, when we wish to evaluate the expectation value of the operator.  Both of these time-evolution operators can be grouped into a single time-ordered operator by formally doubling the field content of the theory to include `$+$' fields associated with the time-evolution of the $|\alpha\rangle$ state and `$-$' fields associated with the evolution of the $\langle\alpha |$ state.  In the interaction picture, since the interacting part of the Hamiltonian $H_I$ is used to evolve the states in the theory, we effectively double the interactions present---for every interaction of the `$+$' fields, there exists a `$-$' field interaction with a coupling of the opposite sign.  Thus the evolution of the expectation value of an operator ${\cal O}$ is given by 
\begin{equation}
\langle\alpha | {\cal O} | \alpha\rangle (\eta) =
{\langle\alpha | T \left\{ {\cal O}_I e^{-i\int_{\eta_0}^0 d\eta'\, [ H_I(\Phi^+) - H_I(\Phi^-) ]} \right\} | \alpha \rangle \over
\langle\alpha | T \left\{ e^{-i\int_{\eta_0}^0 d\eta'\, [ H_I(\Phi^+) - H_I(\Phi^-) ]} \right\} | \alpha \rangle} . 
\label{Omatrix}
\end{equation}
Here $T$ is the time-ordering operator which orders events along the time contour so that the arguments of the $\Phi^-$ fields always occur after, and in the opposite order than, those of the $\Phi^+$ fields.  The subscript in ${\cal O}_I$ indicates that the operator is evaluated in the interaction picture.  The field doubling automatically removes the acausal portion of the matrix element so that although the time-evolution operators integrate to the infinite future, $\eta'=0$ in conformal coordinates, terms involving propagators depending upon $\eta'>\eta$ are cancelled in Eq.~(\ref{Omatrix}).  A more complete description of applying the Schwinger-Keldysh method to a de Sitter background is provided in \cite{fate}.  

For evaluating the perturbative corrections to the power spectrum in an interacting theory, we take the expectation value of the two-point operator ${\cal O}_I = \Phi(\eta,\vec x)\Phi(\eta,\vec y)$ in a truncated $\alpha$-vacuum.  In the interaction picture, the time evolution of operators is produced by the free Hamiltonian so that the fields evolve correctly when the mode functions satisfy the free Klein-Gordon equation~(\ref{KleinGordon}).  In a theory with a cubic interaction, 
\begin{equation}
H_I = \int {d^3\vec x\over H^4\eta^4}\, \left[ J\Phi + {\textstyle{1\over 2}} \delta m^2 \Phi^2 + {\textstyle{1\over 3!}} \lambda \Phi^3 \right] , 
\label{Hint}
\end{equation}
the first non-trivial corrections to the two-point operator appear at one loop order.  The linear and quadratic terms correspond to allowed counterterms; the former is used to cancel tadpole subdiagrams and the latter cancels a logarithmic ultraviolet (UV) divergence in the matrix elements of a true $\alpha$-vacuum \cite{fate}.

\begin{figure}[!tbp]
\includegraphics{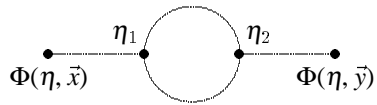}
\caption{The one loop correction to the two-point function in a theory with a cubic interaction used to generate the power spectrum.\label{loopfig}}
\end{figure}

When $J$ is chosen to cancel the tadpole subgraphs, the power spectrum is given to second order in the coupling constant $\lambda$ by 
\begin{eqnarray}
{\cal P}^\alpha(\eta,k)
&=& {k^3\over 2\pi^2} |U_k^\alpha(\eta)|^2 
\nonumber \\
&&- {\delta m^2\over\pi^2} {k^3\over H^4}
\int_{\eta_0}^{\eta} {d\eta_1\over\eta_1^4}\,\, {\rm Im} \bigl\{ 
\bigl[ U_k^\alpha(\eta) U_k^{\alpha*}(\eta_1) \bigr]^2
\bigr\}
\nonumber \\
&&+ {2\lambda^2\over\pi^2} {k^3\over H^8} 
\int_{\eta_0}^{\eta} {d\eta_1\over\eta_1^4}\,\, {\rm Im} \bigl\{ 
U_k^\alpha(\eta) U_k^{\alpha*}(\eta_1) \bigr\}
\nonumber \\
&&\quad\times 
\int_{\eta_0}^{\eta_1} {d\eta_2\over\eta_2^4}\,\, {\rm Im} \bigl\{ 
U_k^\alpha(\eta) U_k^{\alpha*}(\eta_2) L_k^\alpha(\eta_1,\eta_2)
\bigr\}
\nonumber \\
&&+ \cdots 
\label{twopoint}
\end{eqnarray}
where the loop integral is defined by 
\begin{equation}
L_k^\alpha(\eta_1,\eta_2) \equiv
\int^\Lambda_\mu\!\!\! d^3\vec p\, U_p^\alpha(\eta_1)U_p^{\alpha*}(\eta_2)
U_{|\vec p-\vec k|}^\alpha(\eta_1)U_{|\vec p-\vec k|}^{\alpha*}(\eta_2) . 
\label{loopdef}
\end{equation}
Diagrammatically, the order $\lambda^2$ correction to the power spectrum, ${\cal P}^{(2)}(\eta,k)$, is generated by the self-energy graph shown in Fig.~\ref{loopfig}.

The spatial momenta in the loop in Eq.~(\ref{loopdef}) are bounded in the UV by the structure of the truncated $\alpha$-vacuum since we have assumed that the short distance behavior is highly suppressed as in Eq.~(\ref{fullmodes}).  As we shall show, once we have extracted an overall factor of the physical cutoff $\Lambda_{\rm phys}$, the leading contribution to ${\cal P}^{(2)}(\eta,k)$ does not depend sensitively on how we truncate the modes since it arises among terms from the loop whose $p\equiv |\vec p|$ dependent phases cancel.  The integral of these terms add coherently and receive their largest contributions well away from the UV cutoff.  

A massless, minimally coupled scalar theory in a de Sitter background contains infrared (IR) divergences \cite{bunch,dSIR} so we have imposed an IR cutoff, $\mu$, on our loop integrals.  However true inflation is not in a pure de Sitter background and the scalar which provides the source for the perturbations is light but not massless.  We could incorporate a small mass, $m\ll H$, in Eq.~(\ref{Emodes}), and expand in $m/H$ so that no IR cutoff would be necessary.  Yet since we find that the dominant contribution to the loop integral is not strongly sensitive to $\mu$ we shall focus on the UV physics.

With these bounds on the large and short distance behavior, any of the comoving momenta appearing in the integrand of Eq.~(\ref{loopdef}) should always be between the bounds, 
\begin{equation}
\mu \le p \le \Lambda
\qquad
\mu \le |\vec p-\vec k| \le \Lambda . 
\label{bounds}
\end{equation}
Since $\mu$ and $\Lambda$ are the bounds on the comoving momenta, to relate them to the fixed IR and UV cutoffs we must specify how they depend on the conformal times at which they are evaluated.  The limits of the $\eta_1$ and $\eta_2$ integrations in Eq.~(\ref{twopoint}) impose that $\eta_2 \le \eta_1$($<0$).  Therefore, we should choose the UV bound in the loop integration to scale as 
\begin{equation}
\Lambda = {\Lambda_{\rm phys}\over -H\eta_2} 
\equiv - {\tilde\Lambda\over\eta_2}
\label{UVscaling}
\end{equation}
since by the time we have arrived at the $\eta_1$ vertex of the loop, all comoving momenta will have been red-shifted and will therefore still be below $\Lambda$.  Conversely, we would not like any momenta to be redshifted below the IR cutoff by the later time, $\eta_1$, so we choose the IR bound to scale as 
\begin{equation}
\mu = {\mu_{\rm phys}\over -H\eta_1} 
\equiv - {\tilde\mu\over\eta_1} . 
\label{IRscaling}
\end{equation}
For example, when the IR cutoff is larger than the horizon size during inflation, we have that $\tilde\mu\ll 1$.

In addition to establishing the appropriate bounds on the spatial momenta appearing in the one loop correction to the power spectrum, we must determine the limits on the conformal time integrals.  The natural choice for the initial time, $\eta_0$, from which we are evolving the matrix element is at the UV scale since in our model we assume that when the comoving momentum equals this scale that the corresponding mode is that for the alpha vacuum.  Thus we take
\begin{equation}
k\eta_0 = - \tilde\Lambda . 
\label{etaUVbound}
\end{equation}
The lower scale depends upon when the state is assumed to become essentially classical, so that we can neglect the quantum corrections.   Here we shall take the low energy cutoff to be given by the conformal time at which the momentum of the mode $k$ has been redshifted to the size of the IR cutoff,
\begin{equation}
k\eta_\mu = - \tilde\mu . 
\label{etaIRbound}
\end{equation}

After integrating over the spatial loop momenta, the only remaining spatial momentum is $k\equiv |\vec k|$ and it becomes convenient to define dimensionless variables by rescaling the conformal times,
\begin{equation}
x=k\eta, 
\qquad
x_1=k\eta_1, 
\qquad
x_2=k\eta_2 . 
\label{xdefs}
\end{equation}
The leading contribution to the ${\cal O}(\lambda^2)$ correction to the power spectrum comes from the region of the loop integral,
\begin{eqnarray}
L_k^\alpha(x_1,x_2) 
&=& - {2\pi\over k^3} H^4 N_\alpha^4 e^{\alpha+\alpha^*} x_1^2 x_2 \tilde\Lambda 
\label{bigloop} \\
&&\times 
\left[ {\sin(x_1-x_2)\over x_1-x_2} + {\sin(x_1+x_2)\over x_1+x_2} \right]
+ \cdots ,
\nonumber
\end{eqnarray}
and is given by 
\begin{eqnarray}
{\cal P}^{(2)}(\eta,k) 
&=& - {\lambda^2\over\pi} N_\alpha^4 e^{\alpha+\alpha^*} \tilde\Lambda 
\biggl\{ 
\nonumber \\
&&
\int_{\tilde\Lambda-\tilde\mu}^{\tilde\mu} {dx_1\over x_1^2}\, 
\bigl[ (1+xx_1)\sin(x-x_1) 
\nonumber \\
&&\qquad\qquad\quad 
- (x-x_1)\cos(x-x_1) \bigr]
\nonumber \\
&&\times\int_{\tilde\Lambda-\tilde\mu}^{x_1} {dx_2\over x_2^3}\, 
\bigl[ (1+xx_2)\sin(x-x_2) 
\nonumber \\
&&\qquad\qquad\qquad 
- (x-x_2)\cos(x-x_2) \bigr]
\nonumber \\
&&\qquad\qquad 
\times\left[ {\sin(x_1-x_2)\over x_1-x_2} + {\sin(x_1+x_2)\over x_1+x_2} \right]
\nonumber \\
&& 
+\, {\cal O}\bigl(\tilde\Lambda^{-2}\bigr) 
+ {\cal O}\bigl(\tilde\Lambda^{-1}\ln(\tilde\mu)\bigr) \cdots \biggr\} . 
\label{powerlead}
\end{eqnarray}
The momentum dependence of this correction is not of a form that can be cancelled by the counterterms in Eq.~(\ref{Hint}) \cite{fate}.  The dimensionless integral receives its dominant contribution from the region $0.1\alt x_1,x_2\alt 4$ and is not strongly sensitive to the location of the cutoffs.  Integrating numerically, we find that 
\begin{eqnarray}
{\cal I} 
&\equiv& 
\int_{\tilde\Lambda-\tilde\mu}^{\tilde\mu} {dx_1\over x_1^2}\, 
\bigl[ (1+xx_1)\sin(x-x_1) 
\nonumber \\
&&\qquad\qquad\quad 
- (x-x_1)\cos(x-x_1) \bigr]
\nonumber \\
&&\times\int_{\tilde\Lambda-\tilde\mu}^{x_1} {dx_2\over x_2^3}\, 
\bigl[ (1+xx_2)\sin(x-x_2) 
\nonumber \\
&&\qquad\qquad\qquad 
- (x-x_2)\cos(x-x_2) \bigr]
\nonumber \\
&&\qquad\qquad 
\times\left[ {\sin(x_1-x_2)\over x_1-x_2} + {\sin(x_1+x_2)\over x_1+x_2} \right]
\nonumber \\
&&
= - 0.2618 + {\cal O}\bigl( \tilde\Lambda^{-2} \bigr) 
+ {\cal O}\bigl( \tilde\mu^3 \bigr) .
\label{intnumber}
\end{eqnarray}
so that the leading behavior of the corrections is linear in the UV cutoff,
\begin{equation}
{\cal P}^{(2)}(\eta,k) 
= 0.2618 \times {\lambda^2\over\pi} N_\alpha^4 e^{\alpha+\alpha^*} 
{\Lambda_{\rm phys}\over H} 
+ \cdots . \quad
\label{powertwo}
\end{equation}

If we define a dimensionless coupling $\tilde\lambda \equiv \lambda / H$, then the power spectrum for $k\eta\ll 1$ to leading order is 
\begin{eqnarray}
{\cal P}^\alpha(\eta,k) 
&=& {H^2\over 4\pi^2} N_\alpha^2\left| 1 - e^\alpha \right|^2 
\label{powerleading} \\
&&\times\biggl\{ 1 + 1.047 \pi 
{N_\alpha^2 e^{\alpha+\alpha^*}\over \left| 1 - e^\alpha \right|^2}\, 
\tilde\lambda^2 {\Lambda_{\rm phys} \over H} 
+ \cdots \biggr\} . 
\nonumber
\end{eqnarray}

The corrected power spectrum is still independent of $k$ to leading order.  This feature is a consequence of our choice of the matching condition in Eq.~(\ref{fullmodes}) which, as with the full $\alpha$-vacuum itself, was independent of $k$.  If we had instead chosen a $k$-dependent matching condition, as in the modified uncertainty relation of \cite{kempf,gary} or as in the initial conditions chosen in \cite{ulf}, this $k$-dependence would have propagated to the one-loop corrections, so that in general Eq.~(\ref{powertwo}) would vary with $k$.  

Note that the loop correction scales inversely with the Hubble scale $H$ so that in low scale inflation this portion of the correction becomes larger.  In the next section we estimate the size of the small dimensionless coupling $\tilde\lambda$ for some specific models.

\section{The inflaton potential}\label{potential}

While the loop contribution to the power spectrum is enhanced by a factor of $\Lambda_{\rm phys}/H$, the dimensionless coupling constant $\tilde\lambda$ suppresses the total loop term relative to the tree contribution.  The $\alpha$-vacuum enhancement is generic for any cubic self-interaction, but the coupling constant suppression is model dependent.  We briefly consider two common inflationary models to illustrate the size of the coupling and hence the overall size of the loop term.

To determine the size of the coupling we need the COBE normalization constraint~\cite{liddle_lyth},
\begin{equation}
5\times 10^{-4} \approx {V^{3/2}(\phi) \over M_{\rm pl}^3 V'(\phi)}.
\end{equation}
Here a prime denotes differentiation with respect to the field $\phi$:  $V'=\partial V/\partial\phi$.  We also require that the modes observed in the CMB left the horizon about $50$ $e$-folds before the end of inflation, which determines the value of the inflaton zero mode, $\phi$, when these fluctuations left the horizon
\begin{equation}
50 \approx {1 \over M_{\rm pl}^2} \int_{\phi_{end}}^\phi ~
{V(\phi') \over V'(\phi')} d\phi'.
\label{nphi}
\end{equation}
For chaotic inflation, where the potential is dominated by the inflaton terms such as, $V(\phi) = {1\over 3!}H\tilde\lambda\phi^3$, inflation occurs when the inflaton field is much larger than the Planck scale and ends when $\phi_{end}\approx M_{\rm pl}$.  From Eq.~(\ref{nphi}) we have $\phi\approx 20 M_{\rm pl}$ and the COBE normalization gives $\tilde\lambda \approx 10^{-11} M_{\rm pl}/H$.  Use of the Friedmann equation gives the Hubble scale, $H \approx 10^{-4}M_{\rm pl}$ and if we set $\Lambda_{\rm phys}\sim M_{\rm pl}$, we finally arrive at the result we seek,
\begin{equation}
\tilde\lambda^2{\Lambda_{\rm phys} \over H} \approx 10^{-10}.
\end{equation}
Clearly this is a significant suppression.

For a hybrid inflation model, the potential is dominated by a constant term, $V_0$, as long as $\phi$ is above a critical value, $\phi_c$.  $V_0$ must be roughly constant while the modes of interest are leaving the horizon, but contains operators which end inflation at $\phi = \phi_c$.  We use the potential $V(\phi) = V_0 + {1\over 3!}H\tilde\lambda\phi^3$.  If $\phi$ is not fine tuned to be close to $\phi_c$, then Eq.~(\ref{nphi}) gives
\begin{equation}
\phi_c \approx {H \over 10 \tilde\lambda} < \phi.
\end{equation}
Dropping numerical factors close to one, the COBE normalization now gives, $10^{-4} \approx H^2/(\tilde\lambda \phi^2)$.  It is consistent to choose $\phi \approx {\cal O}(\phi_c)$, in which case $\tilde\lambda \approx 10^{-6}$.  The size of the enhancement is difficult to predict without more details of the theory, specifically, the Hubble scale.  However, by requiring that the constant term dominate the potential, $V_0 \gg {1\over 3!}H\tilde\lambda\phi^3$, as it must for this to be a model of hybrid inflation, we arrive at
\begin{equation}
\tilde\lambda^2{\Lambda_{\rm phys} \over H} \gg 10^{-8}.
\end{equation}
For small enough inflationary energy density,
\begin{equation}
V_0 \lesssim (10^{12}{\rm GeV})^4,
\end{equation}
the Hubble parameter is small enough that the loop effects are as large as the tree level effect, or larger.  Of course, if we are to remain in the perturbative regime, we require a suppression from the $\alpha$ dependent factors; the state at the matching scale may only deviate from the Euclidean state by a small amount.

Models for inflation with a cubic self-coupling may not be common, but the enhancement demonstrated in this work is not limited to cubic theories.  The na\"\i ve power counting argument discussed in \cite{fate} indicates that the loop corrections to the 2-point function in a quartic theory should scale as
\begin{equation}
{\cal O}\left(\lambda_4^2 {\Lambda_{\rm phys}^3 \over H^3} \right).
\end{equation}
Although there is again a suppression due to the model dependent coupling constant, the loop enhancement effect is quite general.

In models in which the vacuum state of the inflaton is non-thermal, loop effects can generally be larger than what would be suggested by the counting of coupling constants alone.  In some cases, the model may not even be perturbative for $|\alpha| \sim {\cal O}(1)$ and thereby will constrain the degree to which the vacuum can deviate from the Euclidean, or thermal, vacuum.

\begin{figure}[!tbp]
\includegraphics{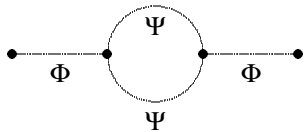}
\caption{Even if the inflaton $\Phi$ is in the Euclidean vacuum, couplings to other scalar fields in a non-thermal vacuum, $\Psi$, will produce large corrections to the power spectrum.\label{trifig}}
\end{figure}

Thus far we have only considered a single scalar field, corresponding to the inflaton.  However, any other fields present will be sensitive to the rapid expansion of the universe and will require a matching condition at the cutoff scale as well.  These fields could also be in a non-thermal vacuum below this scale and their coupling to the inflaton would then provide an additional source for large loop corrections---even if the inflaton itself was in the Euclidean vacuum.  For example, when the inflaton interacts with an additional scalar field $\Psi$ through a trilinear coupling, $g\Phi\Psi^2$, then the power spectrum receives potentially large loop corrections from the diagram shown in Fig.~\ref{trifig}.
When $\Psi$ is in a truncated $\alpha$-vacuum, the product of $\alpha$-vacuum propagators in the loop integral will again produce a $\Lambda_{\rm phys}$ enhancement.

Furthermore, any simple model of inflation, such as those mentioned in this section, must ultimately be embedded in a more fundamental theory with many massive degrees of freedom, such as $\Psi$.  To arrive at an inflationary model, those massive fields must be integrated out leaving an effective Lagrangian for the inflaton.  Given the loop enhancements expected for a generic $\alpha$-vacuum, or in any non-thermal vacuum, this process of integrating out the high energy physics may contain previously unexpected subtleties.

\section{Conclusions}\label{conclude}

In this article we have investigated the perturbative corrections to the power spectrum due to inflaton interactions when the field is in a truncated $\alpha$-vacuum.  In a theory with a cubic interaction with a coupling $\lambda$, the size of the perturbative corrections are of the order $(\lambda^2/H^2)(\Lambda_{\rm phys}/H)$.  The large second factor depends only on the form of the free-field propagators in the loop, and is therefore quite general.  The initial suppressing factor from the coupling constant is constrained more or less restrictively depending upon the specific inflationary model studied.  

For a single self-interacting inflaton in chaotic inflation, slow-roll or observational constraints require a sufficiently small coupling $\lambda$, compared with the Hubble scale $H$, that the overall loop effects in these models will be negligible.  However, in a hybrid model the self-coupling is not nearly as constrained so that loop effects could be potentially observed in the cosmic microwave background radiation, or could even be larger than tree-level effects so that such a theory could not be treated perturbatively.  Moreover, as a result of inverse scaling of the Hubble constant in the $\Lambda_{\rm phys}/H$ factor, such loop corrections are actually more significant in low-scale inflationary theories where tree level corrections in a non-thermal background would be negligible.  

Studying loop corrections to the power spectrum in a non-thermal vacuum provides an additional resource for constraining inflationary models or the vacuum state during inflation, as further precision measurements are made of the CMB.  If deviations from the Euclidean vacuum expectation are observed, it is important to distinguish their possible origins---whether as non-thermal tree \cite{ulf,kempf,gary,garygeneric,lowe,brandenberger,ss} or loop effects or from the dynamics of other fields near or below the Planck scale \cite{rich,nemanja}---to understand what they are telling us about the very early universe.

\begin{acknowledgments}
This work was supported in part by DOE grant DE-FG03-91-ER40682.  We are grateful to Rich Holman for helpful discussions and comments on the manuscript.
\end{acknowledgments}

\end{document}